\documentclass[journal=jctcce,manuscript=article]{achemso}
\usepackage{graphicx} 
\usepackage{dcolumn} 
\usepackage{bm} 
\usepackage[utf8]{inputenc}
\usepackage[T1]{fontenc}
\usepackage{etoolbox}
\usepackage{physics}
\usepackage{hyperref} 
\usepackage[version=4]{mhchem} 
\SectionNumbersOn
 
\title{Angular Momentum Transfer between a Molecular System and a Continuous Circularly Polarized Light Field under the Born-Oppenheimer Framework}

\author{Xuezhi Bian}
\email{xzbian@sas.upenn.edu}
\affiliation{Department of Chemistry, University of Pennsylvania, Philadelphia, Pennsylvania 19104, USA}
\author{Joseph E. Subotnik}
\email{subotnik@sas.upenn.edu}
\affiliation{Department of Chemistry, University of Pennsylvania, Philadelphia, Pennsylvania 19104, USA}
\date{\today}

\begin{document}
\begin{tocentry}
\includegraphics[width=7cm]{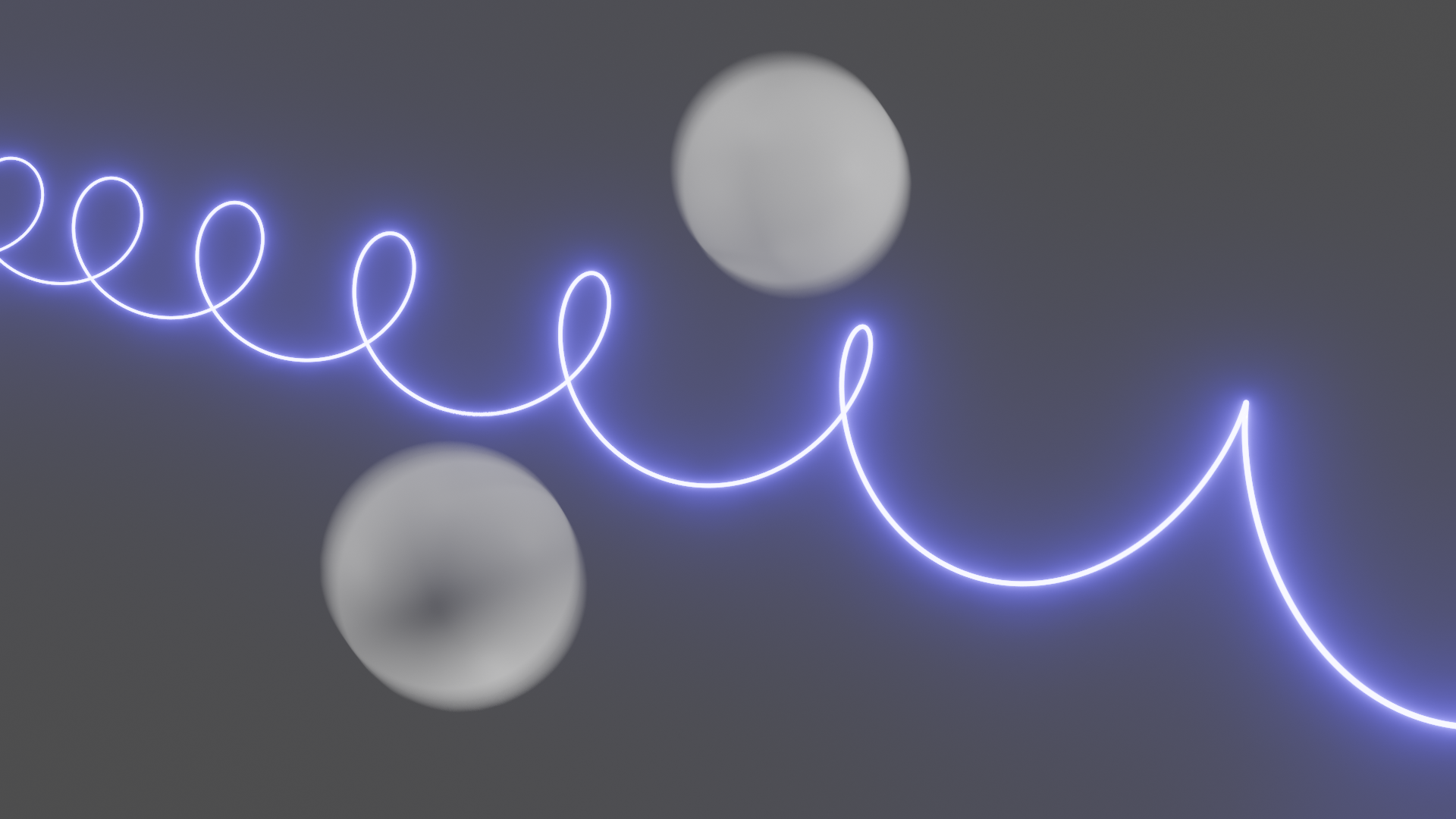}
\end{tocentry}

\begin{abstract}
We demonstrate (both analytically and numerically) total angular momentum conservation for a molecular system subject to circularly polarized light (CPL) field moving along a single  Born-Oppenheimer surface, where all of the angular momentum transfer is embodied in a Berry force. Moreover, we demonstrate that the model Hamiltonian proposed in [J.Chem.Phys. 150, 124101 (2019)] in fact corresponds physically to a homonuclear diatomic in a CPL field.
Our results not only reveal an interesting microscopic mechanism for angular momentum transfer between a molecule and a radiation field, but they also  provide new insight into the nature of novel semiclassical non-adiabatic dynamics methods that conserve 
the total angular momentum  (including, e.g., phase-space surface-hopping methods). 
\end{abstract}
 
\maketitle
\newpage 

\section{Introduction} 
Circularly polarized light (CPL) carries a well-defined spin angular momentum  along its propagation direction. To date, numerous applications have emerged from the interaction between circularly polarized light and the chirality of the matter\cite{Rodger1997,Berova2000}. 
For example, circular dichroism (CD) has long served as a tool to distinguish molecular chiral enantiomers. Moreover, electronic circular dichroism and vibrational circular dichroism have been used extensively to extract structural and conformational information for a wide range of compounds (from small organic molecules \cite{Pescitelli2011} to DNA and proteins\cite{Kelly2005}), and magnetic circular dichroism has been used to study electronic dynamics among different spin states\cite{Stephens1974}. 
All of these different forms of CD spectroscopy rely on the differential absorption of left-hand and right-hand CPL by molecules. 

As observed by Beth almost a century ago \cite{Beth1936}, when matter absorbs circularly polarized light, spin angular momentum is transferred from light to matter, causing a mechanical rotation of the matter, in agreement with the law of total angular momentum conservation. Now within a realistic quantum mechanical calculation, there are many means to maintain angular momentum conservation. 
For instance, let us imagine partitioning the total angular momentum to its nuclear, electronic and photonic components:
\begin{equation}
  {\bm J}_{tot} =  {\bm J}_{n} +   {\bm J}_{e} +  {\bm J}_{ph}.
\end{equation}
Due to the existence of a light-matter interaction, none of these three quantities is a good (i.e. conserved) quantum number.
Thus, when a molecule absorbs a circularly polarized photon and there is a transition from an initial quantum state to a final quantum state, the leftover angular momenta difference of  $\pm \hbar$ can be distributed either to nuclear motion (causing a total rotation of the molecule) or to the electronic motion (causing a ``rotation'' of the electronic distribution). One can ask: Is angular momentum ever exchanged among all three components at once\cite{Franke2017}? How does the angular momentum transfer occur in real time during a photon absorption process? Will the angular momentum transfer affect the intramolecular vibrational relaxation and, ultimately, alter chemical reactivities \cite{Shi2016,Comby2016}?
Addressing these questions will not only enhance our understanding of a fundamental light-matter interaction problem but should also yield insight into new applications, such as designing molecular motors \cite{Galstyan1997} and utilizing circularly polarized light for enantioselectivity in chemical reactions \cite{Kawasaki2005}.

For theoretical and computational chemists, the most straightforward way to answer the questions above is to simulate the real-time quantum dynamics of the light-matter coupled system. However, modeling such processes with the total (nuclei + electrons + external fields) time-dependent Schrodinger equation is extremely challenging due to the computational demand.
Instead, Born-Oppenheimer (BO) theory \cite{Born1985} is routinely employed for studying realistically sized molecules, whereby one first solves the electronic Schrodinger equation with fixed nuclei (with or without an external field) and one then propagates classical nuclear dynamics on one eigensurface.  In this paper, we will investigate how angular momentum transfers between an optical CPL field and a molecule within such a BO framework.  
Our approach is as follows: We will first derive the essential equations of motion (EOM) for  BO dynamics subject to a CPL field, where the role of molecular Berry curvature will be essential both for angular momentum transfer and  total angular momentum conservation.
Second, we will use an {\em ab initio} numerical example to realize the theory just described for a diatomic molecule.
Third and finally, we will extend the discussion of BO dynamics to the realm of non-adiabatic dynamics, and we will show that recent attempts in the literature to develop accurate non-adiabatic surface hopping dynamics are in fact based on an empirical fact that physical trajectories must maintain total angular momentum conservation.

\section{Theory} \label{sec:theory}
Understanding how linear and angular momentum are conserved \cite{Li2022} and how the continuity equation arises semiclassically \cite{Takatsuka2021} for nonadiabatic systems are key challenges that are not extremely well known in the broader chemical physics community.
In a recent set of papers, our research group (and our collaborators) have explored how linear and angular momentum are conserved for field-free systems within the framework of BO dynamics\cite{Bian2023} and/or Ehrenfest dynamics\cite{Tao2023}. Here, we will extend these arguments to demonstrate how the total angular momentum is conserved for a molecule interacting with a CPL field.

Let us start by considering a general Hamiltonian for a molecular system interacting with a radiation field
\begin{equation}
\hat H  =  \hat H_M +  \hat H_R +  \hat H_{MR},
\end{equation}
where $\hat H_M$ represents the molecular Hamiltonian, $\hat H_R$ is the radiation field and $\hat H_{MR}$ is the coupling term between the radiation field and the molecule. 
The molecular Hamiltonian takes the usual form: 
\begin{equation}
    \hat H_M = \hat T_n + \hat H_{e}, 
\end{equation}
where $\hat T_n$ is the nuclear kinetic operator, and $\hat H_e$ is the electronic Hamiltonian. According to BO theory, the first step is to diagonalize $\hat H_{e}$ at a fixed nuclear geometry, 
\begin{equation}
    \hat H_{e} \ket{\phi_j} = E_j \ket{\phi_j}.  
\end{equation}
where $E_j$ is the electronic potential energy and $\ket{\phi_j}$ is the molecular adiabatic basis. 
To simplify our problem, let us imagine that we restrict our discussion to only the two lowest electronic energy surfaces. Then, the Hamiltonian can be represented by a $2\times 2$ matrix in the molecular adiabatic basis, 
\begin{equation} \label{eq:HM}
    \hat H_M = \hat T_n \mathcal{I} + \begin{pmatrix} 
     E_0 & 0 \\ 0 &  E_1
    \end{pmatrix}.
\end{equation}
Within the usual BO approximation, we further assume that the energy gap between the two electronic states is large ($E_1 \gg E_0$) and that all other states are also energetically well separated ($E_n \gg E_1$ for $n>1$). Therefore, we can ignore all of the non-adiabatic couplings and focus exclusively on a two-level system coupled only by a radiation field. 

Now, for the radiation field part, let us consider a monochromatic continuous wave (CW) light with frequency $\omega$. The quantum radiation field Hamiltonian can be written as:
\begin{equation}
    \hat H_R = \hbar \omega \left(\hat a^\dagger \hat a + \frac 1 2\right). 
\end{equation}
where $\hat a$ and $\hat a^\dagger$ are the photon annihilation and creation operators for the given mode. We assume that the frequency of the field is on resonance with (or close to) the electronic excitation of the molecule.   
If the field frequency is in the infrared range, then no electronic excited state will be involved (which is of less interest for this paper). 
Under the dipole approximation, the light-matter coupling can be written as:
\begin{equation} \label{eq:VMR}
    \hat V_{MR} = \begin{pmatrix} 
    0 & \bm \mu_{01} \cdot \bm {\mathcal{\hat E}}   \\ {\bm \mu}_{10} \cdot \bm{\mathcal{\hat E}}  & 0
    \end{pmatrix}, 
\end{equation}
where $\bm \mu_{01} = \bra{\phi_0} \hat {\bm \mu} \ket{\phi_1}$  is transition dipole moment between molecular adiabats $\ket{\phi_0}$ and $\ket{\phi_1}$, ${\bm \epsilon}$ is the polarization vector of the electric field, and 
$\hat {\bm{\mathcal{E}}}  = {\mathcal{E}}_0 ( {\bm \epsilon} \hat a e^{i\omega t} + {\bm \epsilon}^* \hat a^\dagger e^{-i\omega t})$ represents the time-dependent local electric field operator at the molecular position; note that we have here made the long wavelength approximation for $\hat {\bm{\mathcal{E}}}$.
In Eq.~\ref{eq:VMR}, we also assume that the molecular system is non-polar, but a generalization to polar molecules can be trivially done by adding permanent dipole terms to the diagonal elements.    

Since the Hamiltonian in Eqs.~\ref{eq:HM} - \ref{eq:VMR} is periodic in time, according to the rotating wave approximation (which is an approximation to Floquet theory\cite{Horenko2001,Fiedlschuster2016,Schiro2021}), we can rewrite the Hamiltonian in a dressed Born-Oppenheimer basis: $|\tilde {\phi}_0\rangle =  \ket{ \phi_0} \otimes{\ket{N+1}}$ and $|\tilde {\phi}_1\rangle  =  \ket{\phi_1} \otimes{\ket{N}}$, where $\ket{N+1}$ and $\ket{N}$ are the photon wavefunctions with $N+1$ and $N$ photons in the particle number representation\cite{Nitzan2006}. 
\begin{equation}  \label{eq:HF}
    \hat H =  \hat T_{n}\mathcal{I} + \left(N + \frac 1 2\right)\hbar\omega  \mathcal{I} + 
    \begin{pmatrix} 
    E_0 +\hbar \omega & \bm \mu_{01} \cdot  \bm{\epsilon}\mathcal{E}_0    \\ {\bm \mu}_{10} \cdot \bm{\epsilon^*}\mathcal{E}_0   & E_1
    \end{pmatrix}.
\end{equation}
Note that the physical picture here is simple:  $|\tilde\phi_0\rangle$ represents the molecule  in the ground state with $N+1$ photons and $|\tilde\phi_1\rangle$ represents the molecule in the excited state with $N$ photons. 

At this point, we have obtained an effective Hamiltonian for a molecule in radiation field. As we stated before, propagating full quantum dynamics with Hamiltonian in Eq.~\ref{eq:HF} for complex molecules is very difficult. 
We will now consider the simplest case -- propagating classical Born-Oppenheimer molecular dynamics (BOMD) on a single electronic surface. 
To start, we diagonalize the Hamiltonian in the molecular adiabatic basis in Eq.~\ref{eq:HF}:
\begin{equation}
    \hat H \ket{\psi_j} =   E_i^a \ket{\psi_j},
\end{equation}
where $\ket{\psi_j}$ is the true ``adiabatic basis'' for the light-matter Hamiltonian and $ E_j^a $ is the corresponding eigenenergy. 

Then, the nuclear wave packet $\chi_j$ on the $j$-th surface will follow an effective Schrodinger equation:
\begin{equation} \label{eq:Heff}
    i\hbar \frac {\partial} {\partial t} \ket{\chi_j} = \left( \frac {(\hat {\bm P} - \bm A_{jj})^2} {2M} + E_j^a \right) \ket{\chi_j}, 
\end{equation}
where $\hat {\bm P} = -i\hbar {\bm\nabla}_{\bm R}$ is the nuclear momentum operator, ${\bm d}_{jj} = \bra{\psi_j} \bm \nabla_{\bm R} \ket{\psi_j}$ is the derivative coupling and ${\bm A}_{jj} = i\hbar {\bm  d}_{jj} $ is the nuclear Berry connection. \cite{Bohm2003} Note that the nuclear Berry connection term stems from the nuclear coordinates dependency of the eigenbasis $\ket{\psi_j}$  and depends only very indirectly on the vector potential of the radiation field (which itself is independent of any internal nuclear coordinate and moreover, by the long wavelength approximation, does not depend on the molecular center of mass either). By defining the nuclear kinetic momentum operator as $\hat {\bm \pi} = \hat {\bm P} - {\bm A_{jj}}$ and utilizing 
the Heisenberg equation of motion, we can derive EOMs from Eq.~\ref{eq:Heff}:
\begin{equation} \label{eq:EOMR}
    \dot {\bm R} =  \frac { \bm \pi} M, 
\end{equation}
\begin{equation} \label{eq:EOMpi}
    \dot {\bm \pi} =  -\bm \nabla_{\bm R} E_j + {\bm \Omega}_{jj} \cdot \frac {\bm \pi } M, 
\end{equation}
where 
\begin{equation} \label{eq:BC}
   \bm \Omega_{jj} = \bm \nabla \times \bm A_{jj} 
\end{equation}
is the nuclear Berry curvature. 
In Eq.~\ref{eq:EOMR} and Eq.~\ref{eq:EOMpi}, we have replaced the quantum operators with their classical counterparts. The first term in Eq.~\ref{eq:EOMpi} represents the usual BO force on a single energy surface and the second term is the Berry force caused by the nuclear gauge field. 
Eqs.~\ref{eq:EOMR} and ~\ref{eq:EOMpi} are the working equations for  single surface BOMD with a light field.  

Let us now turn to the angular momentum in such a situation. For a classical molecular trajectory, defining nuclear angular momentum is trivial: 
\begin{equation} \label{eq:Jnuc}
    \bm J_{n} = \bm R \times \bm \pi. 
\end{equation} 
For the electronic angular momentum, we define
\begin{equation}\label{eq:Jele}
    {\bm J}_{e} = \bra{\psi_j(\bm R)} \hat {\bm J}_{e} \ket{\psi_j(\bm R)}. 
\end{equation}
where $\hat {\bm J}_{e}$ is the electronic angular momentum operator that operates in the electronic Hilbert space. 
If the photon mode of the radiation field is circularly polarized,  
we can denote the single photon basis as $\ket{L}$ for the left-hand circularly polarized mode and $\ket{R}$ for the left-hand circularly polarized mode. Then, if we suppose the field is propagating along the $\hat z$ direction, a single-photon angular momentum operator can be defined as $\hat {J}^z_{ph} = \hbar(\ket{L}\bra{L} - \ket{R}\bra{R})$. Within the photon Fock space, where one can find $N$ photons in a left-handed or right-handed polarized mode, we have:
\begin{equation}
    \hat {J}^z_{ph} \ket{N_L} = N\hbar  \ket{N_L} , \quad  \hat {J}^z_{ph} \ket{N_R} = -N\hbar \ket{N_R}.
\end{equation}
Thus,  if the photon mode is left polarized, when $\hat J^z_{ph}$ is projected into our light-dressed two-state BO basis, the photon angular momentum operator reads:
\begin{equation}\label{eq:Jph}
    \hat {J}^z_{ph, \rm projected} = \hbar \begin{pmatrix}
        N+1  & 0 \\ 
        0  & N 
    \end{pmatrix}.
\end{equation}
For simplicity, we will omit the ``projected'' label and use $\hat{{\bm J}}_{ph}$ to denote the phonon angular momentum operator in the dressed BO basis throughout the rest of the paper. 

Finally, for a single classical BO dynamics trajectory, the total angular momentum can be expressed as:
\begin{equation}
    {\bm J}_{tot} = {\bm J}_{n} + \bra{\psi_j (\bm R)} \hat {\bm J}_{e} + \hat {\bm J}_{ph} \ket{\psi_j (\bm R)} 
\end{equation}
with the corresponding quantities defined above. 

As guaranteed by the Noether's theorem for an isolated system of light and molecules, the total angular momentum must be a conserved quantity. Let us now prove such a conservation law within the BO framework.  We begin by considering the time evolution of the total angular momentum along a BOMD trajectory:
\begin{equation}
    \frac {d{\bm J}_{tot}} {dt}  =    \frac {d{\bm J}_{n}} {dt} + \frac {d\bra{\psi_j (\bm R)} \hat {\bm J}_{e} + \hat {\bm J}_{ph} \ket{\psi_j (\bm R)}} {dt}.
\end{equation}

As shown in Ref.~\citenum{Littlejohn2023}, for angular momentum conservation, BO theory requires the electronic eigenbasis follow a well-defined phase convention that satisfies translational and rotational invariance\cite{Littlejohn2023}:
\begin{equation}
(\hat {\bm P}_{n} +   \hat {\bm P}_{e}) \ket{\psi_j} = 0,
\end{equation}
\begin{equation}
(\hat {\bm J}_{n} +   \hat {\bm J}_{e}) \ket{\psi_j} = 0.
\end{equation}
For a more detailed discussion of this phase convention, see Ref. ~\citenum{Littlejohn2023} . Here we will generalize this phase convention to the case with a radiation field:
\begin{equation} \label{eq:tphase}
(\hat {\bm P}_{n} +   \hat {\bm P}_{e} + \hat {\bm P}_{ph} ) \ket{\psi_j} = 0,
\end{equation}
\begin{equation} \label{eq:rphase}
(\hat {\bm J}_{n} +   \hat {\bm J}_{e} + \hat {\bm J}_{ph} ) \ket{\psi_j} = 0.
\end{equation}
Following the phase convention as defined by Eq.~\ref{eq:tphase} and Eq.~\ref{eq:rphase},  we have
\begin{equation}
\begin{aligned} \label{eq:dJ}
    \frac {d{\bm J}_{tot}} {dt}  &=  \frac {d{\bm J}_{n}} {dt} - \frac {d\bra{\psi_j (\bm R)} \hat {\bm J}_{n}  \ket{\psi_j (\bm R)}} {dt} \\
    & =  \frac {d{\bm J}_{n}} {dt} + i\hbar {\bm R} \times  \frac {d\bra{\psi_j (\bm R)} \nabla_{\bm R} \ket{\psi_j (\bm R)}} {dt} \\
    &= \frac {d{\bm J}_{n}} {dt} + i\hbar \sum_k {\bm R} \times  \left( \bra{ \frac {d{\psi_j }} {dt} } \ket{\psi_k}\bra{\psi_{k}}\nabla_{\bm R} \ket{\psi_j} + \bra{\psi_{j}}\nabla_{\bm R} \ket{\psi_k} \bra{ \frac {d{\psi_k }} {dt} } \ket{\psi_j} \right) \\ 
    &=  \frac {d{\bm J}_{n}} {dt} + \frac {2} {\hbar} \sum_k  {\bm R}  \times {\rm Im} \left( \left( \dot {\bm R} \cdot \bm A_{jk} \right) \bm A_{kj} \right).
\end{aligned}
\end{equation}
For calculating Berry curvature, the following identity is easy to derive from Eq.~\ref{eq:BC} above:
\begin{equation} \label{eq:BCindex}
    \Omega^{I\alpha J\beta}_{jj} = -\frac 2 \hbar {\rm Im} \sum_{k\neq j} A_{jk}^{I\alpha} A_{kj}^{J\beta} . 
\end{equation}
where $I,J$ are nuclear labels and $\alpha, \beta$ are Cartesian indices. Plugging Eq.~\ref{eq:BCindex} into Eq.~\ref{eq:dJ}, we find 
\begin{equation} \label{eq:dJtot}
 \frac {d{\bm J}_{tot}} {dt} =   \frac {d{\bm J}_{n}} {dt} - {\bm R} \times \left( {\bm \Omega}_{jj}   \cdot \frac {\bm \pi} M \right). 
\end{equation}
According to the EOM in Eqs.~\ref{eq:EOMR} and \ref{eq:EOMpi}, the change of nuclear angular momentum can be expressed as:
\begin{equation} \label{eq:dJnuc}
    \frac {d{\bm J}_{n}} {dt} = {\bm R} \times \left( -\nabla_{\bm R} E_j + {\bm \Omega}_{jj}   \cdot \frac {\bm \pi} M \right).
\end{equation}
Clearly, the second term in Eq.~\ref{eq:dJtot} (representing the change of electronic and photonic angular momentum)  is exactly canceled by the change in nuclear angular momentum arising from the Berry force (the second term in Eq.~\ref{eq:dJnuc}). Moreover, according to the rotational invariance of BO energy surfaces, the first term Eq.~\ref{eq:dJnuc} will be zero. Thus, in the end, we have shown that with the proper phase conventions, BOMD with a Berry force conserves the total angular momentum:
\begin{equation} \label{eq:dJtot0}
    \frac {d{\bm J}_{tot}} {dt} = 0. 
\end{equation}

One cannot emphasize enough that, for single surface BO molecular dynamics in a CPL field, the total angular momentum is conserved if and only if one includes the nuclear Berry force.   In particular, transfer of angular momentum between photonic DOFs and molecular DOFs is clearly realized by the nuclear Berry force. Note that when the radiation field is linearly polarized, the Hamiltonian in Eq.~\ref{eq:HF} will be completely real-valued, the nuclear Berry force will vanish,  and no angular momentum transfer will occur between molecule and radiation field. 

\section{Results} \label{sec:result}
\subsection{Numerical Demonstration of Total Angular momentum Conservation in BOMD} \label{subsec:numerics}

We will now employ an {\em ab initio} simulation to study the theory above, demonstrate total angular momentum conservation, and quantify angular momentum transfer for a molecule exposed to a circularly polarized radiation field. 

We consider a ${\rm H}_2$ molecule with the H-H bond aligned along the $\hat x$ direction.
Such diatomic systems are often considered to be a platform for artificial control of non-adiabatic dynamics through light-matter interactions (e.g., light-induced conical intersections).\cite{Demekhin2013,Natan2016,Badanko2018,Bouakline2018,Farag2021,Carrasco2022}
We assume the CPL field is propagating along the $\hat z$ direction and is left-hand polarized ($\hat \epsilon = \hat x + i\hat y$) with frequency $\hbar\omega = 0.5$ Hartree. To fit our model,
we calculated the energies of the lowest two singlet eigenstate $S_0$, $S_1$ for the ${\rm H}_2$ molecule at the restricted Hartree-Fock and configuration interaction singles (RHF/CIS) level of theory with the 6-311G* basis set as a function of the bond length ($R$). We further computed the transition dipole moments between these two states as we all the derivative of transition dipoles with respect to nuclear coordinates; the latter are needed for calculating the nuclear Berry curvature. 
All electronic structure calculations were performed in a developmental version of Q-Chem 6.0\cite{Epifanovsky2021}.

\begin{figure}[ht]
\includegraphics[width=\textwidth]{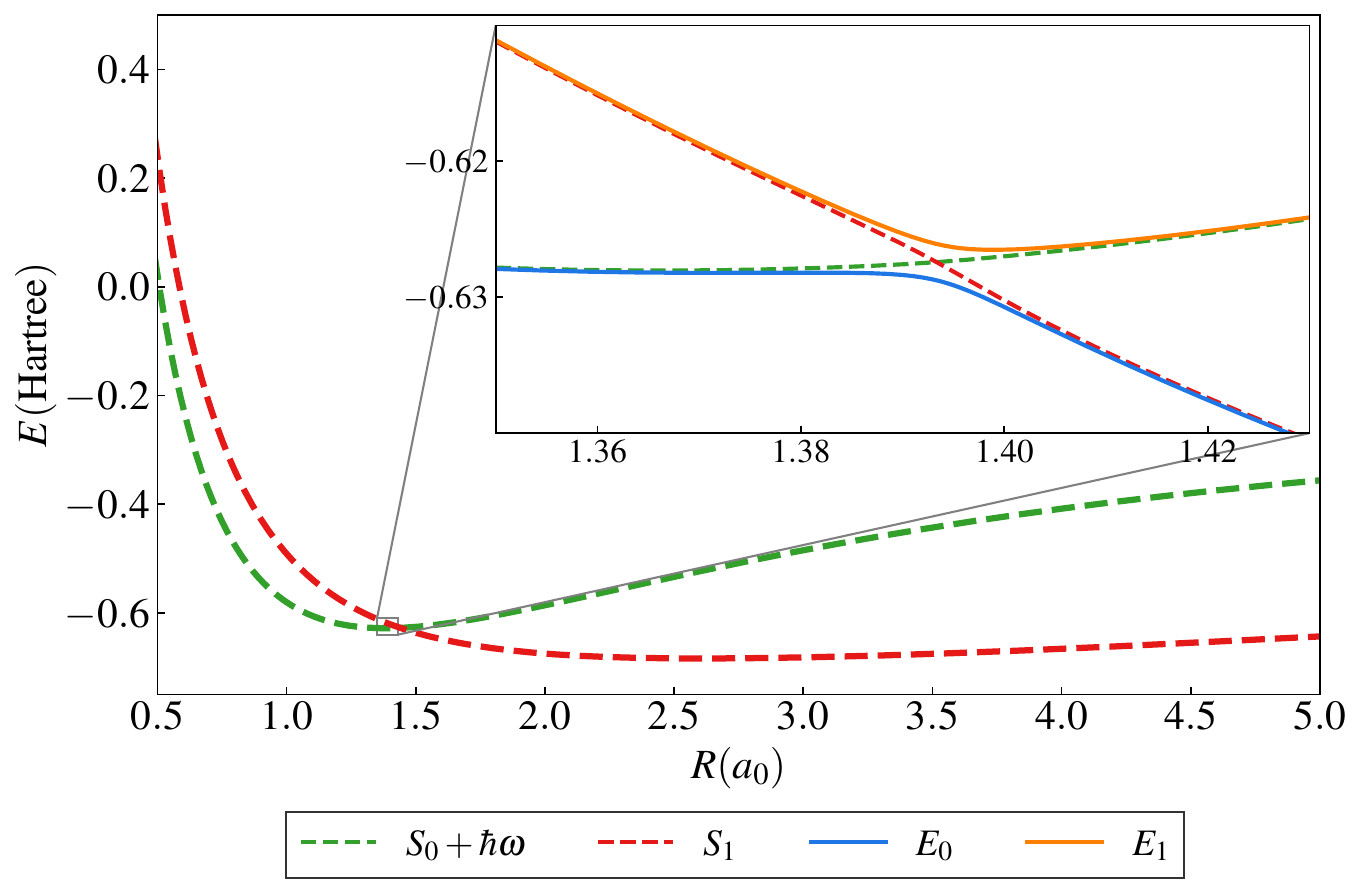} 
\caption{\label{fig:PES} Potential energy surfaces of the molecular ground state (plus a photon,red) and first excited state (without a photon, green) of a ${\rm H}_2$ molecule as a function of the H-H bond length $R$. 
The enlarged section of the figure zooms in on the splitting and plots the diabatic (green and red) and adiabatic (blue and yellow) energies in the dressed representation. Notably, there is a light-induced avoided crossing observed at approximately $1.39 a_0$.} 
\end{figure}

In Fig.~\ref{fig:PES}, 
we plot the energy profile for the bare ${\rm H}_2$ molecule and the total system (molecule + radiation field) eigenstate $E_0$, $E_1$ as a function of H-H bond length over the range $R \in \left[0.5a_0,5.0a_0\right]$.
From the zoomed in figure in Fig.~\ref{fig:PES}, one can clearly identify a light-induced avoided crossing at around $1.39 a_0$.
Note that the process of a $\rm H_2 $ molecule moving through the light-induced avoided crossing from the left (ground state minima) to the right adiabatically on the ground state corresponds to a photo-dissociation reaction. Note that the early time dynamics should be accurate even though we use a RHF potential (that cannot capture bond-breaking asymptotically).

\begin{figure}[ht]
\includegraphics[width=\textwidth]{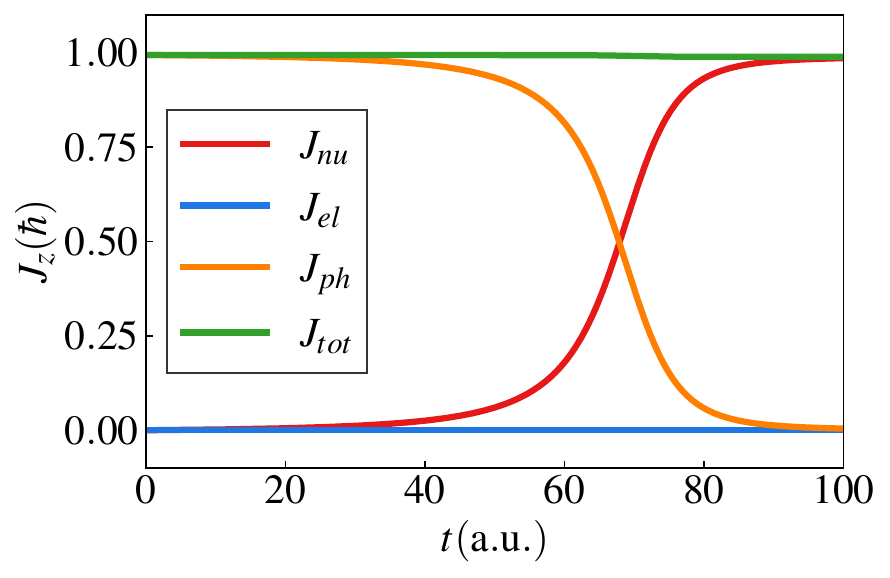} 
\caption{ \label{fig:BOMD}
The nuclear, electronic and photonic contributions to the real-time total angular momentum for a classical BOMD trajectory traversing the light-induced avoided crossing with nuclear Berry force. The total angular momentum (green) is conserved , as $\hbar$ angular momentum is transferred from the radiation field (yellow) to nuclei (red).
The electronic angular momentum (blue) remains zero during the propagation (see text).
} 
\end{figure}

Next, we initialized the ${\rm H}_2$  molecule at $R_0 = 1.35 a_0$ on the ground state  $E_0$ with momentum $P_0 = 1.5 $ a.u. (an initial stretch along the direction of vibration). 
Then, we propagated Born-Oppenheimer molecular dynamics with EOMs in Eqs.~\ref{eq:EOMR}-~\ref{eq:EOMpi} for the $\rm H_2$ molecule on the ground adiabatic surface. 
For our initial condition, the $\rm H_2$ molecule has enough energy to move through the light-induced avoided crossing and reach the dissociation region. 
We calculated the nuclear, electronic and photonic angular momentum for a single BOMD trajectory in real time according to Eqs.~\ref{eq:Jnuc}-~\ref{eq:Jph}. The results for the angular momentum in the $\hat z$ direction for the nuclear, electron, and photonic  DOFs (plus the total)   are plotted in Fig.~\ref{fig:BOMD}.

As shown in Fig.~\ref{fig:BOMD}, the total angular momentum is always conserved during the propagation with Berry force,  where the nuclear angular momentum changed from zero initially to exactly $\hbar$ while the photon angular momentum changed from $\hbar$ to 0. However, the electronic angular component of the angular momentum remains zero throughout the propagation. This state of affairs appears counter-intuitive because, when an optical photon is absorbed by a molecule, we  expect the photonic energy to transfer to the electronic DOFs as the molecular wavefunction moves along an electronic excited state, whereas our results show that the angular momentum is transferred entirely and directly to nuclei. 
To that end, one must recognize that this observation of direct transfer of angular momentum is quite a coarse-grained view that depends sensitively on the number of state we include in our model. In particular, note that the model above makes many approximations, of which two are essential to recognize here:
(i) Within our modified BO framework, because we include only two electronic states (ground and excited) and each state is dressed by a different number of photons (one or zero),  the electronic angular momentum operator (Eq.~\ref{eq:Jele})  is diagonal in our dressed BO basis $|\tilde \phi_j \rangle$ (defined below Eq.~\ref{eq:Jedress}),
\begin{equation}\label{eq:Jedress}
    \langle \tilde \phi_j | \hat {\bm J}_{e} | \tilde \phi_k\rangle = \delta_{jk} \bra{\phi_j} \hat {\bm J}_{e} \ket{\phi_k}.
\end{equation}
(ii) Although the operator $J_{e}$ should include spin angular momentum in principle, here we deal with singlet states, so that there is no spin component. Moreover,  the molecular states $\ket{\phi_1}$ and $\ket{\phi_2}$ are chosen to be real-valued so that $\bra{\phi_k} \hat {\bm J}_e \ket{\phi_k} = 0$.
Thus, in fact by construction, $J_e$ can never be nonzero inside of our limited two-state model above.
That being said, we emphasize that that one {\em would} find non-zero electronic angular momentum if one did not make the approximations above, i.e. if one were to work with spin degrees of freedom. Moreover, if one did not impose the rotating wave approximation or if one went beyond the dipole approximation for the light-matter coupling term (e.g., up to the magnetic order -- the magnetic dipole interaction will couple electronic states with different spin), one should also expect non-zero transient electronic angular momentum. 

\section{Discussion: Mapping a Diatomic Molecule in a CPL field to a Classic Model Hamiltonian}

In this section, 
we wish to point out that the model above of a homonuclear diatomic molecule in a CPL field is closely related to a model Hamiltonian that has been used to study non-adiabatic semiclassical dynamics with a complex-valued Hamiltonian\cite{Miao2019}. 

Let us rewrite the Hamiltonian in Eq.~\ref{eq:HF} as:
\begin{equation}
\begin{aligned}
    \hat H &= (\hat T_n + N\hbar\omega) \mathcal{I} + \begin{pmatrix} 
    E_0 +\hbar \omega & \bm \mu_{01} \cdot  \bm{\epsilon}\mathcal{E}_0    \\ {\bm \mu}_{10} \cdot \bm{\epsilon^*}\mathcal{E}_0   & E_1
    \end{pmatrix}\\
    &= \label{eq:gaohan} (\hat T_n  + \bar E) \mathcal{I}  +  C\begin{pmatrix}
       -\cos\theta & \sin\theta e^{i\varphi} \\
        \sin\theta e^{-i\varphi} & \cos\theta 
    \end{pmatrix}
    \\
    & = (\hat T_n  + \bar E) \mathcal{I} + \hat{H}_{e}(\theta,\varphi)
\end{aligned}
\end{equation} 
In Eq.~\ref{eq:gaohan}, we define:
\begin{equation} \label{eq:theta}
    \theta = \arctan(\frac {2 \abs{{\bm \mu}_{01}  \cdot   \bm{\mathcal{E}}}} {E_0 + \hbar \omega - E_1}), 
\end{equation}
\begin{equation}\label{eq:phi}
    \varphi =  \arg  \left(   {\bm \mu}_{01}  \cdot   \bm{\mathcal{E}} \right), 
\end{equation}
\begin{equation}\label{eq:A}
    C  =  \frac 1 {\sqrt{ \abs{{\bm \mu}_{01}  \cdot   \bm{\mathcal{E}}}^2 + \frac 1 4 (E_0 + \hbar\omega - E_1)^2}},
\end{equation}
\begin{equation}
    \bar E = N\hbar\omega + \frac 1 2 (E_0 + E_1 + \hbar \omega).
\end{equation}
Suppose initially the molecule is placed in the $xy$-plane with no out-of-plane motion while light is propagating along the $\hat z$ direction. Clearly, the molecule will remain in  the $xy$-plane.
Since the direction of the diatomic molecular transition dipole is always parallel to the direction between that two different atoms, according to Eq.~\ref{eq:phi}, the parameter $\varphi$ is simply the polar coordinate $\varphi$ in the $xy$-plane where we work in the lab frame. Furthermore, the parameters $\theta$, $C$ and $\bar E$ are  functions of only one variable -- the bond length $R$. 

To define the nuclear kinetic energy operator in terms of $(\theta, \varphi)$ coordinates, we note that for any system in two dimensions expressed in polar coordinates $(R,\varphi)$, the standard form for the Laplacian is
\begin{equation}
    \hat T_n = \frac {\hat P_R^2}{2M} + \frac {\hat P_\varphi^2} {2MR^2}, 
\end{equation}
where $\hat P_R = -i\hbar(\nabla_R + (2R)^{-1})$ and $\hat P_\varphi = -i\hbar\nabla_\varphi$ are the canonical momentum operators that satisfy the canonical commutation relation with their conjugate position variables $\hat R$ and $\hat \varphi$.
Then, under a coordinate transformation $R \to \theta(R), \varphi \to \varphi$, the new kinetic energy operator can be expressed in terms of the coordinates $\theta$ and $\varphi$ as follows:
\begin{equation} \label{eq:Tn}
    \hat T_n = \frac {1}{2M}\left(-\frac {i\hbar \nabla_\theta} {f'(\theta)}  - \frac {i\hbar} {2f(\theta)} \right)^2  + \frac {\hat P_\varphi^2} {2M f^2(\theta)}, 
\end{equation}
where we have defined $R = f(\theta)$. 
According to Eq.~\ref{eq:Tn}, it reasonable to define the canonical momentum operator for the variable $\theta$ as $\hat P_\theta = -i\hbar(\nabla_\theta + \frac {f'} {2f}) $.
One can easily prove that $\hat P_\theta$ is self-adjoint and it satisfies the commutation relation $\left[\hat \theta, \hat P_\theta \right] = i\hbar$.

At this point, we have fully constructed the Hamiltonian in Eq.~\ref{eq:gaohan} in terms of $\theta$ and $\varphi$. The final step is then to diagonalize the corresponding electronic Hamiltonian $\hat H_e (\theta,\varphi)$ in Eq.~\ref{eq:gaohan}. 
The relevant eigenstates are:
\begin{equation} \label{eq:wf}
    \psi_0 = 
    \begin{pmatrix} 
    -\cos \frac {\theta} 2 e^{i\varphi}   \\ \sin \frac {\theta} 2 
    \end{pmatrix}, \quad 
    \psi_1 = 
    \begin{pmatrix} 
    \sin \frac {\theta} 2 e^{i\varphi}   \\ \cos \frac {\theta} 2 
    \end{pmatrix}.
\end{equation}
In such a basis (that depends on $\theta$ and $\varphi$), one must replace the derivatives inside of the momentum operator $\hat P_\theta$ and $\hat P_\varphi$ by their covariant derivatives, such that the nuclear kinetic energy operator becomes:
\begin{equation}
    \hat T_n = \frac { (\hat P_\theta - A_\theta)^2 }{2M (f'(\theta))^2}   + \frac {(\hat P_\varphi - A_\varphi)^2} {2M f^2(\theta)}, 
\end{equation}
where $A_\theta$ and $A_\varphi$ are the corresponding Berry connections. 
According to Eq.~\ref{eq:wf}, the ground state nuclear Berry connection and Berry curvature can be calculated as:
\begin{equation}
    A_\theta = \bra{\psi_0} \nabla_\theta\ket{\psi_0} = 0, \quad A_\varphi =  \bra{\psi_0} \nabla_\varphi\ket{\psi_0} =  i\hbar \cos^2 \frac {\theta} 2, 
\end{equation}
\begin{equation}
    \Omega_{\varphi\theta} = \nabla_\theta A_\varphi - \nabla_\varphi A_\theta = \frac \hbar 2 \sin\theta = -\Omega_{\theta\varphi}.
\end{equation}

We can further define the kinetic momentum operators as before:
\begin{equation}
\hat \pi_\theta  = \hat P_\theta - A_\theta, \quad \hat \pi_\varphi  = \hat P_\varphi - A_\varphi. 
\end{equation}
Thus, in  a compact form, the total Hamiltonian becomes:
\begin{equation}
\label{eq:gaohanad}
    \hat H = \frac {\hat \pi^2_\theta }{2M (f'(\theta))^2}   + \frac {\hat \pi^2_\varphi} {2M f^2(\theta)} + 
    \begin{pmatrix}
        E_0 & 0 \\ 
        0 & E_1
    \end{pmatrix},
\end{equation}
where $E_0 = \bar E - C$ and   $E_1 = \bar E + C$.

\subsection{Born-Oppenheimer Dynamics}
If we apply the Heisenberg equations of motion to the Hamiltonian in Eq.~\ref{eq:gaohanad} and further make a  classical approximation, the EOMs for the position and momentum coordinates become:
\begin{equation}
    \dot \theta =  \frac {\pi_\theta}{M(f'(\theta))^2},
\end{equation}
\begin{equation}
    \dot \varphi = \frac {\pi_\varphi} {M f^2(\theta)},
\end{equation}
\begin{equation}
    \dot \pi_\theta = -\nabla_\theta E - \frac {\pi_\varphi^2 {f'(\theta)}} {M  f^3(\theta)}    + \Omega_{\theta\varphi} \dot \varphi,
\end{equation}
\begin{equation}
    \dot \pi_\varphi = \Omega_{\varphi\theta} \dot \theta.
\end{equation}

Notice that the EOM for $\pi_\varphi$ is extremely simple: there is only a Berry force contribution. 
At the risk of redundancy, let us now demonstrate how the total angular momentum conserved for our diatomic molecule  in terms of the coordinates $\theta$ and $\varphi$ using Eq.~\ref{eq:dJtot0}. To that end, for the diatomic molecular system,  we note that $\pi_\varphi$  represents the nuclear angular momentum and its EOM is given by:
\begin{equation} \label{eq:dJnucz}
    \frac {dJ_n^z} {dt} = \dot \pi_{\varphi} = \frac {\hbar } {2 } \dot \theta \sin\theta
\end{equation}
With the analytical wavefunction in Eq.~\ref{eq:wf} and the definition in Eq.~\ref{eq:Jph}, the EOM for the photon angular momentum is given by:
\begin{equation} \label{eq:dJph}
  \frac {d\langle J_{ph}\rangle}{dt} =  \frac {d\bra{\psi_0(\theta,\varphi)} \hat J_{ph}^z \ket{\psi_0(\theta,\varphi)}} {dt} = -\frac {\hbar} {2} \dot \theta \sin\theta.
\end{equation}
Comparing Eq.~\ref{eq:dJnucz} and Eq.~\ref{eq:dJph}, the change in photon angular momentum cancels exactly with the change in nuclear angular momentum at any time and leads to total angular momentum conservation. 

We can also show that the angular momentum molecule transferred from the photon is quantized. 
Suppose the molecule is initialized away from the light-induced avoided crossing ($E_0 + \hbar\omega \ll E_1$), so that we can define $\theta = 0$ according to Eq.~\ref{eq:theta}. 
After propagating to the far other side of the crossing ($E_0 + \hbar\omega \gg E_1$), we find $\theta = \pi$. For this process whereby a  trajectory passes through the avoided crossing, i.e., $\theta = 0 \to \theta = \pi$, the total change of nuclear angular momentum is given by
\begin{equation}
    \Delta J_n = \Delta \pi_{\varphi} = \int_{t=0}^{t=\infty} \dot \pi_\varphi dt = \int_{t=0}^{t=\infty} \Omega_{\varphi\theta} \dot \theta dt = \int_{\theta=0}^{\theta=\pi} \Omega_{\varphi\theta} d\theta = \hbar. 
\end{equation} 

For single surface BOMD, our proof above reaches the same conclusions as were found above in Sec.~\ref{sec:theory}: the nuclear Berry force is essential for  total angular momentum conservation in single surface dynamics. 
However, the single surface dynamics is valid only in the adiabatic limit, i.e., when the light-matter interaction is strong enough. The elephant in the room is now clear: what about non-adiabatic dynamics?

\subsection{Non-adiabatic Dynamics and Surface Hopping}

The Hamiltonian in Eq.~\ref{eq:gaohan} is intimately related to a model that our research group and others have worked on over the past five years \cite{Miao2019,Miao2020,Wu2021,Wu2022,Krotz2023,Zhou2023}. As described in Ref.~\citenum{Miao2019},
this model incorporates two electronic states and two Cartesian nuclear DOFs $x$ and $y$:
\begin{equation} \label{eq:gaohan3}
    \hat H' = \left( \frac {P_x^2}{2M} + \frac {P_y^2}{2M}   \right) \mathcal{I}  +  \hat H_e
\end{equation}

The electronic potential energy matrix $\hat H_e$ is expressed in the general form in Eq.~\ref{eq:gaohan}, where the nuclear dependency of the parameters $\theta$ and $\varphi$ are given as:
\begin{equation} \label{eq:gaohan2}
    \theta = \frac \pi 2 (\erf(Bx) + 1), \quad \varphi = Wy. 
\end{equation}
The parameter $\theta$ is chosen to be a function of $x$, so that there is a crossing in the diabatic representation  as $\theta$ goes from 0 to $\pi$ ($x$ goes from $-\infty$ to $\infty$); $B$ is a constant that controls the sharpness of the crossing. 
The other parameter $\varphi$ is defined as a function of $y$ where $W$ is also a constant. Note that the dependence of the Hamiltonian on $y$ appears only in the phase of the complex couplings; thus the potential energy surface is completely flat in the $y$ direction. 
The model in Ref. \citenum{Miao2019} ignores the variables $C$ and $\bar E$ in Eq.~\ref{eq:gaohan}
(of effectively just chooses them to be constants); this assumption simplifies the model even further such that the adiabatic potential energy surfaces become flat in the $x$-direction as well. 

In Refs.~\citenum{Miao2019,Miao2020,Wu2021,Wu2022}, our research group studied the dynamics of the Hamiltonian in Eqs.~\ref{eq:gaohan3}-\ref{eq:gaohan2} both rigorously (using exact quantum dynamics ) and semiclassically (using surface hopping dynamics).
As far as exact dynamics are concerned, 
we observed an interesting feature: Although the potential energy surfaces are completely flat along both the $x$ and $y$ directions, 
as the nuclear wave packet passes through the avoided crossing adiabatically, its momentum in the $y$ direction increases (or decreases depending on the wave packet's direction of propagation and initial adiabat) exactly by a factor of  $\hbar W$ (where $W$ is defined in Eq.~\ref{eq:gaohan2}).  
By constrast, if the wave packet passes through the avoided crossing nonadiabatically and transitions to a different adiabat, its momentum in the $y$ direction is unchanged.    
This model Hamiltonian  is very similar to Eq.~\ref{eq:gaohan} where parameterized a diatomic molecule in a CPL field. 
The observation that the momentum changes in the $y$ direction for the Hamiltonian in Eqs.~\ref{eq:gaohan3}-~\ref{eq:gaohan2} matches exactly with the observation that the angular momentum transfers in the diatomic system described by Eq.~\ref{eq:gaohan}. These Hamiltonians are quite similar if one sets $W=1$ and ignores the difference in the kinetic energies. 
In this language, passing through the avoided crossing along the same adiabat (for $\hat H'$) gets mapped to the diatomic molecule absorbing (or emitting) a circularly polarized photon.

Now, modeling the dynamics above using a semiclassical surface hopping ansatz \cite{Tully1990} has proven challenging, but ultimately taught us a great deal about semiclassical dynamics that are often hidden with purely electrostatic Hamiltonians \cite{Bian2021}.  A brief reminder about surface hopping is now appropriate.
Within a surface hopping scheme, one propagates  classical nuclear dynamics for a swarm of trajectories on a single electronic energy surface. At the same time, one integrates the electronic Schrodinger equation for each trajectory and allows every trajectory to stochastically hop to other electronic surfaces based on the change of electronic amplitudes. When a trajectory hops, one rescales its momentum along the direction of the derivative coupling \cite{Pechukas1969,Herman1984,Tully1990,Kapral1999} between the initial state and the final state to conserve the total energy. 
The FSSH method has been used in a wide range of problems including photo-relaxation\cite{Nelson2020,Mai2020}, electron transfer \cite{Jones1998} and light-driven dynamics\cite{Fiedlschuster2017,Mignolet2019,Zhou2020} and it is known for often giving the correct branching ratios and equilibrium distribution\cite{Barbatti2011,Wang2016,Mai2018}.   

Unfortunately, surface hopping suffers from several drawbacks. 
As pointed out in Ref.~\citenum{Miao2019}, one simple obstacle is that there is no clear means to deal with a complex-valued Hamiltonian -- which often arises if the spin DOFs are taken into account or if external electromagnetic fields are considered.
In such a case, the electronic wavefunction can be complex-valued if the electronic Hamiltonian is complex-valued, and therefore so are the derivative couplings which makes the direction for momentum rescaling quite ambiguous -- one cannot rescale the nuclear momentum along a complex-valued direction.

\subsubsection{Momentum Rescaling Between Adiabats Tied Directly to Conservation of Angular Momentum}


After extensive benchmarking for the Hamiltonian in Eqs.~\ref{eq:gaohan3}-\ref{eq:gaohan2}, in Ref.~\citenum{Wu2021}, we  found that the best momentum rescaling direction within a FSSH protocol is  to rescale momentum  so that every trajectory can leave the crossing region with the ``correct'' momentum on the $y$ direction. 
For example, consider a trajectory initialized on the adiabat $\ket{\psi_0}$ at $t=0$, for which the trajectory hops only once at $t=t'$ and leaves the crossing region on the excited adiabat $\ket{\psi_1}$ at $t = \infty$. For such a trajectory, the change of momentum in the $\varphi$ (or $y$) direction $\Delta \pi_\varphi$  during a hop is given by solving:
\begin{equation} \label{eq:rescale}
    \pi_\varphi[0,0\to t'] + \Delta  \pi_\varphi + \pi_\varphi[1,t'\to \infty] = 0,
\end{equation}
where $\pi_\varphi[0,0\to t']$ denotes the momentum change in the $\varphi$ (or $y$) direction for the trajectory propagated on adiabat $\ket{\psi_0}$ with Eq.~\ref{eq:EOMpi} from $t=0$ to $t = t'$. 
Having picked out the component of momentum in the $\varphi$ direction, the momentum change in the $\theta$ (or $x$) direction $\Delta \pi_\theta$ is then solved by insisting upon total energy conservation (which is standard\cite{Tully1990}). Note that this approach chooses a momentum rescaling direction dynamically insofar as the direction depends on both trajectory's position and momentum.

Let us now map the above rescaling rules to the case of a diatomic molecule in a CPL field. Since $\pi_\varphi$ is essentially the nuclear angular momentum of the diatomic molecule, solving Eq.~\ref{eq:rescale} enforces total angular momentum conservation (molecular + photonic) during hopping events in surface hopping methods. 
Moreover, given the fact that single surface BO dynamics with nuclear Berry force always conserve total angular momentum, the rescaling algorithm in Ref.~\citenum{Wu2021} in fact conserves the total angular momentum for any given trajectory at all times!
Thus, although unbeknownst to us at the time, 
the algorithm  proposed in Ref.~\citenum{Wu2021} was simply 
a momentum rescaling direction that conserved angular momentum; and perhaps not surprisingly, enforcing such a conservation rule led to strong results.

\subsubsection{Angular Momentum Conservation Achieved By Running on Phase-Space Adiabats}

Beyond the ``best'' FSSH algorithm described above\cite{Wu2021},
for the complex-valued Hamiltonian in Eqs.~\ref{eq:gaohan3}-\ref{eq:gaohan2},
we have found that the best results are in fact achieved by running a phase-space surface hopping (PSSH) algorithm. A few words are now appropriate as far as differentiating  PSSH from FSSH, explaining why PSSH performs so well (including within the strong adiabatic coupling limit\cite{Wu2022}), and in particular why PSSH automatically satisfies total angular momentum conservation  for the Hamiltonian in Eq.~\ref{eq:gaohan}. 

Let us now briefly review the PSSH approach.
PSSH dynamics dictate that nuclei are propagated not along adiabatic surfaces (parameterized by nuclear position) but rather along phase space adiabatic surfaces (parameterized by nuclear position and momentum)~\cite{Wu2022}. To develop such surfaces,  we introduce a pseudo-diabatic transformation that cancels the complex phases in Eq.~\ref{eq:gaohan}: 
\begin{equation} \label{eq:PD}
    |\tilde \phi_0 \rangle = \ket{\phi_0}, \quad  |\tilde \phi_1 \rangle = e^{-i\varphi} \ket{\phi_1},
\end{equation}
so that the total Hamiltonian in a pseudo-diabatic basis becomes:
\begin{equation} \label{eq:HPD}
    \begin{aligned}
         \hat H^{PS} &=  \left( \frac {(\hat P_\theta - {\tilde A}_\theta)^2 }{2M (f'(\theta))^2}   + \frac {(\hat P_\varphi - {\tilde A}_\varphi)^2} {2M f^2(\theta)} + \bar E \right) \mathcal{I} + C 
    \begin{pmatrix}
        -\cos\theta & \sin\theta \\
        \sin\theta &  \cos\theta 
     \end{pmatrix} \\
     &\equiv  \left( \frac {\hat P_\theta^2 }{2M (f'(\theta))^2}   + \frac {\hat P_\varphi ^2} {2M f^2(\theta)} \right)\mathcal{I}   + H_{e}^{PS}(\theta)
    \end{aligned}
\end{equation}
where
\begin{equation}\label{eq:HPS}
\hat H_{e}^{PS}(\theta,P_{\varphi}) = 
     C 
    \begin{pmatrix}
        -\cos\theta & \sin\theta \\
        \sin\theta & \cos\theta + \frac {- 2\hbar P_\varphi + \hbar^2} {2MCf^2(\theta)}
     \end{pmatrix}
\end{equation}
Above, we have defined the Berry connection (in a pseudo-diabatic basis) as  $\tilde {A}_\theta =0 $ and $\tilde {A}_\varphi = -\hbar  |\tilde \phi_1 \rangle  \langle \tilde \phi_1|$. 
By diagonalizing the phase-space Hamiltonian $\hat H_e^{PS}$ in Eq.~\ref{eq:HPS}, one generates ``phase-space adiabats'',
\begin{equation} \label{eq:Eps}
    \hat H^{PS}\ket{\psi_k^{PS}} = E_k^{PS} \ket{\psi_k^{PS}},
\end{equation}
where the phase-space adiabatic energies $E_k^{PS}$ and eigenbasis $\ket{\psi_k^{PS}}$ depend on both nuclear position and momentum.

At this point, we propagate classical nuclear dynamics on a single phase-space adiabat following Hamilton's equations of motion for canonical variables $\bm R = (\theta,\phi)$ and $\bm P = (P_{\theta}, P_{\phi})$:
\begin{equation} \label{eq:Hamilton}
    \frac{d\bm R} {dt} =  -\nabla_{\bm P} E_{k}^{PS}, \quad 
    \frac{d\bm P} {dt} =  -\nabla_{\bm R} E_{k}^{PS};
\end{equation}
In the spirit of standard FSSH, we allow trajectories to hop between different phase-space adiabats according to the electronic Schrodinger equation under the phase-space adiabatic basis.   

The transformation in Eq.~\ref{eq:HPD} removes the dependence on $\varphi$ from the electronic Hamiltonian $\hat H_e^{PS}$ and instead leads to a non-zero $\tilde A_\varphi$ in the kinetic energy operator. Under such a transformation and by analogy to the discussion in Sec.~\ref{sec:theory}, the canonical (angular) momentum $P_\varphi$ effectively represents system's total angular momentum. In principle, one can  define the kinetic angular momentum as $\pi_\varphi^k = P_\varphi - \bra{\psi_k^{PS}} \tilde {A_\varphi} \ket{\psi_k^{PS}}$ (which represents the nuclear angular momentum on phase-space adiabat $k$), but in the context of total angular momentum conservation, one would like to find $ \dot P_{\varphi} = 0$.  To prove this conservation law, for the two-state Hamiltonian in Eq.~\ref{eq:HPS}, we can compute the phase-space adiabats analytically:
\begin{equation} \label{eq:Epm}
    E_{\pm} =   \frac {\hat P_\theta^2 }{2M (f'(\theta))^2}   + \frac {\hat P_\varphi ^2} {2M f^2(\theta)}  + \frac {g} {2}  \pm \frac 1 2 \sqrt{4C^2 +  {g^2} + 4Cg\cos \theta} 
\end{equation}
where 
\begin{equation} \label{eq:g}
    g = \frac {-2\hbar P_\varphi + \hbar^2} {2M f^2(\theta)}.  
\end{equation}
Plugging Eqs.~\ref{eq:Epm} and \ref{eq:g} into Eq.~\ref{eq:Hamilton}, we find that indeed:
\begin{equation}
\label{angfinal}
    \frac {d P_\varphi} {dt} = -\frac {\partial E_\pm} {\partial \varphi} = 0 
\end{equation}
Eq. \ref{angfinal} states that there is total angular momentum conservation for a trajectory moving along a single phase-space adiabatic surface, a feature which we have found makes PSSH so accurate and powerful here. 

As a side note, one can also notice that, after the transformation in Eq.~\ref{eq:HPD}, the Hamiltonian and the derivative couplings between phase-space adiabats become real-valued and do not depend on the angle $\varphi$ explicitly. 
Thus, during a hop between phase-space adiabats, the momentum of the trajectory can be safely rescaled along the direction of the phase-space adiabatic derivative coupling
and the total angular momentum is conserved within the PSSH method.

\section{Conclusion and Outlook} 
In summary, we have demonstrated how angular momentum transfers between molecular systems and a CW-CPL light field under the BO framework.  Although standard (classical) BO dynamics without nuclear Berry force do not the conserve angular momentum of a system with coupled nuclear and electronic motion\cite{Bian2023},
we have shown that a single surface approach with a nuclear Berry force offers a meaningful (and very practical) approach  to study angular momentum transfer processes even for system experiencing strong light-matter interactions, an approach will is practical for both model and {\em  ab initio} BOMD simulations. Moreover, by mapping the dynamics of a diatomic molecule in CPL field to a model Hamiltonian that is now well studied in the literature\cite{Wu2022}, our analysis makes clear that, on a practical level,  dynamical algorithms that conserve the total angular momentum consistently outperform those algorithms that do not achieve such conservation.

Looking forward, this work opens many potential avenues for future study within the BO framework. 
First, beyond BO dynamics, we have noted that PSSH methods can also enforce  angular momentum conservation\cite{Wu2024} and should be extremely useful for studying illuminated systems undergoing nonadiabatic motion.  For systems {\em without} light-matter interaction,  our group has recently suggested that such Hamiltonians can in fact be generated using electron translational and rotational factors\cite{Qiu2024,Tao2024}, an ansatz which is also applicable for {\em ab initio} calculations but will need to be rigorously tested in the future. 
Turning now to systems {\em with} light-matter interactions, an extremely interesting future direction for research will be to see if one can merge a Floquet approach\cite{Bajo2012}
with a standard electrostatic phase-space electronic Hamiltonian by adding the $\bm \mu \cdot \bm {\mathcal{E}}$ coupling appropriately. 
A good first means to check the validity of such a calculation  will be to check for both energy conservation\cite{Subramanian2020, Heindl2021} {\em and} angular momentum conservation of the total nuclear plus electronic plus photonic system for a circularly polarized CW source (which was not addressed in Ref. \cite{Zhou2023}).
If possible, one would also like to extend any dynamics to 
an arbitrary field, e.g. a enveloped light pulse interacting with molecules. In such a case, the angular momentum of the light field is not quantized, one will need to propagate the dynamics of the radiation field explicitly, but the total angular momentum of a closed systems must still be conserved. 


Second, apart from the spin angular momentum of circularly polarized light, helical modes of a radiation field such as the Laguerre-Gaussian beam carries quantized orbital angular momentum (OAM)\cite{Allen1992}. 
OAM transfer between light and atomic matter and electrons has been observed in various experiments\cite{Piccirillo2001,Babiker2002,Picon2010,Mondal2014}. However, to our knowledge, no one has studied OAM transfer between realistic molecules and OAM light theoretically within a BO framework (with meaningful nuclear dynamics). Performing such simulations remains another exciting direction for future investigation.

Third,  a long-standing problem in chemical dynamics has been to achieve enantiomer selectivity using circularly polarized light \cite{Huck1996}.  Within the context of the present simulations, we note that upon exposure to a CW-CPL field, some of the energy imparted to the molecule must lead to nuclear rotational motion (in order to maintain angular momentum conservation), and there slightly less energy must deposited into vibrational motion (relative to illumination by linearly polarization). This difference in the intramolecular energy distribution is likely quite small (because of the small quantity of angular momentum), but if non-trivial spin statistics can emerge from such illumination, there is the possibility that one can generate new chiral-specific chemical dynamics. Extending the present light-matter calculations to included spin degrees of freedom will be essential in the future, especially if we seek to connect with CISS experiments \cite{Naaman2012,Peer2015,Kim2021}.

Fourth and finally, with recent advances in chiral polaritonics, it has been proposed that one may be able to create ultra-strong circularly polarized light-matter interactions in a chiral cavity. In the future, if possible, one would like to use BOMD and semiclassical non-adiabatic dynamics to study the resulting polariton and investigate how 
angular momentum is distributed and transferred between  molecules inside a chiral cavity that display collective behavior\cite{Schafer2023,Riso2023}

\begin{acknowledgement}
This work has been supported by the U.S. Department of Energy, Office of Science, Office of Basic Energy Sciences, under Award no. DE-SC0019397 (J.E.S.). X.B. thanks Yanze Wu, Clàudia Climent  and Zeyu Zhou for helpful discussions.
\end{acknowledgement}

\bibliography{main}
\end{document}